\documentclass[preprint,showpacs,preprintnumbers,amsmath,amssymb]{revtex4}
\newcommand{\non}{\nonumber\\}
\newcommand{\bea}{\begin{eqnarray}}
\newcommand{\eea}{\end{eqnarray}}
\usepackage{epsfig}
\usepackage{graphicx}
\usepackage{dcolumn}
\usepackage{bm}
\let\jnfont=\rm
\def\NPB#1,{{\jnfont Nucl.\ Phys.\ B }{\bf #1},}
\def\PLB#1,{{\jnfont Phys.\ Lett.\ B }{\bf #1},}
\def\EPJC#1,{{\jnfont Eur.\ Phys.\ Jour.\ C }{\bf #1},}
\def\PRD#1,{{\jnfont Phys.\ Rev.\ D }{\bf #1},}
\def\PRL#1,{{\jnfont Phys.\ Rev.\ Lett.\ }{\bf #1},}
\def\MPLA#1,{{\jnfont Mod.\ Phys.\ Lett.\ A }{\bf #1},}
\def\JPG#1,{{\jnfont J.\ Phys.\ G}{\bf #1},}
\def\CTP#1,{{\jnfont Commun.\ Theor.\ Phys.\ }{\bf #1},}
\def\JHEP#1,{{\jnfont JHEP \ }{\bf #1},}
\def\NPPS#1,{{\jnfont Nucl.\ Phys.\ Proc.\ Suppl.\ }{\bf #1},}

\def\oversim#1#2{\lower0.5ex\vbox{\baselineskip0pt\lineskip0pt
  \lineskiplimit0pt\everycr{}\tabskip0pt
  \halign{$\mathsurround0pt #1\hfil##\hfil$\crcr #2\crcr\sim\crcr}}}
\begin{document}

\preprint{\parbox{1.2in}{\noindent  arXiv: 0801.1169}}

\title{\ \\[3mm]  B-meson Dileptonic Decays in NMSSM with a Light CP-odd Higgs Boson}

\author{\ \\[2mm] Zhaoxia Heng$^a$,  Robert J. Oakes$^b$, Wenyu Wang$^c$, Zhaohua Xiong$^d$,
                  Jin Min Yang$^c$ \\ ~}

\affiliation{$^a$ Physics Department, Henan Normal University,
              Xinxiang 453007, China\\
 $^b$ Department of Physics and Astronomy, Northwestern University,
              Evanston, IL 60208, USA\\
 $^c$ Institute of Theoretical Physics, Academia Sinica,
              Beijing 100080, China \\
 $^d$ Institute of Theoretical Physics, College of Applied Sciences,
              Beijing University of Technology, Beijing 100020, China
 \vspace*{1.0cm}}

\begin{abstract}
In the next-to-minimal supersymmetric model (NMSSM)
a light CP-odd Higgs boson is so far allowed by current experiments,
which, together with a large $\tan\beta$, may greatly enhance the rare dileptonic
decays $B\to X_s\ell^+\ell^-$ and $B_s\to\ell^+\ell^-\gamma$.
We examine these decays paying special attention to the new operator
allowed by the light CP-odd Higgs boson.
We find that in the parameter space allowed by current experiments like LEP II and
$b\to s \gamma$,
the branching ratios of these rare decays can be greatly enhanced and thus the existing
experimental data on $B\to X_s \mu^+ \mu^-$ can further stringently constrain the parameter
space (especially the region with a super-light CP-odd Higgs boson and large $\tan\beta$).
In the surviving parameter space we give the predictions for other dileptonic decay
branching ratios and also show the results for the forward-backward asymmetry.
\end{abstract}


\maketitle

\section{Introduction}

Recently some non-minimal supersymmetric models such as the
next-to-minimal supersymmetric model (NMSSM) have attracted much
attention \cite{reviewMSSM} since these models can solve the
$\mu$-problem and alleviate the little hierarchy. In the NMSSM,
for example, the $\mu$-term in the superpotential is forbidden by
imposing a discrete $Z_3$ symmetry and instead it is generated
through the coupling between the two Higgs doublets and a newly
introduced gauge singlet scalar which develops a vacuum
expectation value of the order of the SUSY breaking scale.  In
this way, the $\mu$ parameter at the weak scale can be naturally
explained. The NMSSM can ameliorate the little hierarchy by either
tuning the parameters to enhance the theoretical upper bound for
the mass of the lightest CP-even Higgs boson or relaxing the LEP2
bound of 114 GeV through allowing for a light CP-odd Higgs boson
($A_1$) with mass below $2m_b$ \cite{hierarchy}.

It is interesting to note that in the NMSSM the lightness of such
a CP-odd Higgs boson can be naturally predicted in the enlarged
parameter space, and is also allowed by the LEP II data
\cite{reviewMSSM}. This light Higgs boson can not only alleviate
the little hierarchy, but also can help to explain  the observed
anomaly in the decay $\Sigma^+\to p\mu^+\mu^-$ \cite{He-HyperCP}.
On the other hand, if a Higgs boson is indeed so light, its
effects in some low energy processes may be sizable and thus are
necessary to check \cite{hiller,Ellwanger}. For a super light
$A_1$ the decay $b\to A_1 s$ is open and an analysis has been
performed in \cite{hiller} (note that as analysed in
\cite{hiller}, a small CP-odd Higgs mass is only protected from
RGE-effects in the limit of large $\tan\beta$). In this work we
consider the full possible mass range of $A_1$(heavy,
intermediately heavy and light) and check the NMSSM effects in the
rare B meson dileptonic decays $B\to X_s\ell^+\ell^-$ and
$B_s\to\ell^+\ell^-\gamma$ \cite{hurth}.

These rare dileptonic decays are induced by the flavor-changing
neutral-current (FCNC) $b\to s$ transition and are of special
interest due to their relative cleanness and high sensitivity to
new physics. In the Standard Model (SM) such FCNC processes are
suppressed and have very small branching ratios \cite{Ghinculov04}
but can be greatly enhanced in some new physics models
\cite{Dai97,Huang99,Aliev97,Iltan00,Xiong01,other}. Since
experimental data on $B\to X_s \mu^+ \mu^-$ is available and the
future LHCb or super B factory will further scrutinize B meson
decays, these dileptonic decays serve as a good probe of new
physics.

In supersymmetric models these dileptonic decays can be drastically enhanced by
large $\tan\beta$ since both the $b\to s$ transition loops (such as the charged
Higgsino loops) and the Higgs couplings in the Higgs-propagated diagrams are
proportional to $\tan\beta$. It has been shown (see e.g. \cite{Xiong01}) that in the minimal
supersymmetric model (MSSM) great enhancements are possible for these decays.
In the context of NMSSM, in addition to the $\tan\beta$ enhancement, the presence
of a light CP-odd Higgs boson could further enhance these dileptonic decays.
Thus, the parameter space,  especially the region with a super light CP-odd Higgs
boson and a very large $\tan\beta$, is constrained by the existing data
on $B\to X_s \mu^+ \mu^-$.
In our analysis we will examine the NMSSM effects in these dileptonic decays
by scanning over the parameter space allowed by the LEPII experiments and
the data on $b\to s\gamma$. We will show the $2\sigma$ constraints from
$B\to X_s \mu^+ \mu^-$ on the parameter space and then give the
predictions for other dileptonic decay branching ratios and forward-backward asymmetry.

A key point in our calculations is the presence of a new operator
due to the light CP-odd Higgs boson. In contrast to
the MSSM, where all Higgs bosons and sparticles are so heavy that they
can be integrated out at the weak scale, the light CP-odd Higgs boson
$A_1$ in the NMSSM cannot be integrated out at the weak scale and thus a new
operator $\mathcal{O}_A$ describing the interaction $A_1 b\bar s$ must be
treated carefully.

This work is organized as follows.
In Sec. \ref{sec:model} a brief description of the NMSSM is presented.
In Sec. \ref{Sec:CACQ} we calculate the Wilson coefficients  paying
special attention to the new operator $\mathcal{O}_A$.
In Sec. \ref{sec:decay} we evaluate the NMSSM effects on the dileptonic
decay branching ratios and
the forward-backward (FB) asymmetry, and present some numerical results.
The conclusion is given in Sec. \ref{sec:summary} and
analytical expressions from our calculations are presented
in the Appendix.

\section{A brief description of NMSSM}
\label{sec:model}

In the NMSSM a singlet Higgs superfield $\hat{S}$ is introduced.
A discrete $\mathbb{Z}_3$ symmetry is imposed and thus
only the cubic and trilinear terms are allowed in
the superpotential. The Higgs terms in the superpotential
are then given by
\begin{eqnarray}\label{superpotential}
\lambda\hat{S}\hat{H_u}\cdot\hat{H_d}+\frac{\kappa}{3}\hat{S}^3 \, .
\end{eqnarray}
Note that there is no explicit $\mu$-term and an effective $\mu$-parameter is generated
when the scalar component ($S$) of $\hat{S}$ develops a vacuum expectation value $s/\sqrt{2}$:
\begin{eqnarray}\label{mu-eff}
\mu_{eff}=\lambda \langle S\rangle = \frac{s \lambda }{\sqrt{2}} \, .
\end{eqnarray}
The corresponding soft SUSY-breaking terms are given by
\begin{eqnarray}\label{soft-term}
A_\lambda \lambda S H_u\cdot H_d+\frac{A_\kappa}{3}\kappa S^3 \, .
\end{eqnarray}
The scalar Higgs potential is then given by
\begin{eqnarray}\label{potential}
V_F &=& |\lambda H_u\cdot H_d+ \kappa S^2|^2
     + |\lambda S|^2 \left(|H_d|^2+|H_u|^2 \right)\, , \\
V_D &=&\frac{g_{2}^2}{2} \left( |H_d|^2|H_u|^2-|H_u \cdot H_d|^2 \right)
      +\frac{G^2}{8} \left( |H_d|^2-|H_u|^2\right)^2\, , \\
V_{\rm soft}&=&m_{d}^{2}|H_d|^2 + m_{u}^{2}|H_u|^2
            + m_s^{2}|S|^2
           + \left( A_\lambda \lambda S H_u\cdot H_d
           + \frac{\kappa}{3} A_{\kappa} S^3 + h.c. \right)\, ,
\end{eqnarray}
where  $G^2=g_1^2+g_2^2$ with $g_1$ and $g_2$ being the coupling constants
of $U_Y(1)$ and $SU_L(2)$, respectively.

The scalar fields are expanded as follows:
\begin{eqnarray}
H_d  = \left ( \begin{array}{c}
             \frac{1}{\sqrt{2}} \left( v_d + \phi_d + i \varphi_d \right) \\
             H_d^- \end{array} \right) \, ,
H_u  = \left ( \begin{array}{c} H_u^+ \\
       \frac{1}{\sqrt{2}} \left( v_u + \phi_u + i \varphi_u \right)
        \end{array} \right)  \, ,
S  = \frac{1}{\sqrt{2}} \left( s + \sigma + i \xi \right)  \, .
\end{eqnarray}
The mass eigenstates can be obtained by unitary rotations
\begin{eqnarray}
\left( \begin{array}{c} H_1 \\ H_2\\ H_3\end{array} \right)
= U^H \left( \begin{array}{c} \phi_d \\ \phi_u \\ \sigma \end{array} \right),~
\left(\begin{array}{c} A_1 \\ A_2 \\ G_0\end{array} \right)
= U^A \left(\begin{array}{c}\varphi_d \\ \varphi_u\\ \xi \end{array} \right),~
 \left(\begin{array}{c}G^+\\ H^+ \end{array} \right)
=U \left(\begin{array}{c} H_d^+ \\ H_u^+\end{array}  \right),
\end{eqnarray}
where $H_{1,2,3}$ and $A_{1,2}$ are respectively the CP-even and CP-odd
neutral Higgs bosons,  $G^0$ and $G^+$ are Goldstone bosons, and $H^+$ is
the charged Higgs boson.
It is clear that the charged Higgs sector is the same as in the
MSSM, while the neutral Higgs sector contains one more CP-even
and one more CP-odd Higgs boson.
$U^A$ and $U^H$ are given by
\begin{eqnarray}
U^A=\left( \begin{array}{ccc}
    C_{\theta_A} S_\beta & C_{\theta_A} C_\beta  &  S_{\theta_A} \\
    -S_{\theta_A}S_\beta & -S_{\theta_A}C_\beta  &  C_{\theta_A}\\
    -C_\beta &   S_\beta &  0
     \end{array} \right) , ~~
U^H = \left( \begin{array}{ccc}
\frac{1}{\tan \beta}(C_{\theta_H} -\frac{ v}{s}\delta_+S_{\theta_H})
 &C_{\theta_H} & -S_{\theta_H}  \\
\frac{1}{\tan \beta}(S_{\theta_H} +\frac{v}{s} \delta_+C_{\theta_H})&S_{\theta_H} &
C_{\theta_H} \\
    1 \ & \frac{-1}{\tan \beta} \; & \frac{-v}{s \tan \beta}
\delta_+  \end{array} \right),
\end{eqnarray}
where $C_X=\cos X$ and $S_X=\sin X$ ($X=\theta_A,\theta_H$).
The mixing angles are given by \cite{Miller:2003ay}
\begin{eqnarray}
\theta_A=\frac{\pi}{2}+\frac{v}{s\tan\beta}\delta_-+
\mathcal{O}\left(\frac{1}{\tan^2\beta}\right), ~~
\tan (2\theta_H)=\frac{2 \lambda^2 v s}{2\kappa^2 s^2-m_Z^2}
\end{eqnarray}
with $v\simeq 246$ GeV and
\begin{eqnarray}
\delta_\mp& =& \frac{\sqrt{2}A_\lambda \mp  2 \kappa s}{\sqrt{2}A_\lambda + \kappa s}.
\end{eqnarray}
The Lagrangian of the Higgs couplings to
quarks for a large $\tan \beta$ are given by
\begin{eqnarray}
\label{eq:A1bb}
&& {\cal{L}}_{A_i \bar d d}= -i \frac{g_2 m_d}{2 m_W}  \left(
\frac{v}{s} \delta_- A_1, \tan \beta A_2 \right) \bar d \gamma_5 d, \\
&& {\cal{L}}_{A_i \bar u u}= -i \frac{g_2 m_u}{2 m_W} \frac{1}{\tan \beta}
\left(\frac{ \delta_-v }{s \tan \beta} A_1 , A_2 \right)  \bar u \gamma_5 u ,\\
&& {\cal{L}}_{(H_1,H_2,H_3) \bar d d}=  -\frac{g_2 m_d}{2 m_W}
    \left( ( C_\theta -\frac{ v}{s} \delta_+ S_\theta) H_1,
    (S_\theta +\frac{v}{s} \delta_+ C_\theta) H_2, \tan \beta H_3 \right) \bar d  d ,\\
&&    \mathcal{L}_{(H_1,H_2,H_3) \bar u u}=  -\frac{g_2 m_u}{2 m_W}
\left(C_\theta H_1, S_\theta H_2,- \frac{H_3 }{\tan \beta} \right) \bar u u .
\label{Lhhqq}
\end{eqnarray}
Here one can see that
the coupling of the neutral CP-odd Higgs $A_i$ with up-type quarks
are suppressed by a large $\tan\beta$ and thus can be neglected in
the large $\tan\beta$ limit.

Since one more Higgs superfield $\hat S$ is introduced in the NMSSM,
we have a new neutral Higgsino $\psi_S$. So the neutralino sector
is composed of $U_Y(1)$ gaugino $\lambda^1$, $SU_L(2)$ gaugino $\lambda^2$,
and the Higgsinos  $\psi_{H_d}^1$, $\psi_{H_u}^2$ and  $\psi_S$.
The corresponding mass terms are given by
\begin{eqnarray}
{\cal L}_{m_\chi^0} & = & i\frac{1}{2}v(g_2\lambda^2-g_1\lambda^1)
    (\cos\beta\psi_{H_d}^1-\sin\beta\psi_{H_u}^2)
  - \frac{1}{2} M_2 \lambda^2 \lambda^2
  -\frac{1}{2} M_1 \lambda^1 \lambda^1- \frac{1}{\sqrt{2}}\lambda s \psi_{H_d}^1 \psi_{H_u}^2\non
&& -\frac{1}{\sqrt{2}}v\lambda\left(\cos\beta \psi_{H_u}^2 \psi_S
   +\sin\beta\psi_{H_d}^1 \psi_S\right)
   +\frac{1}{\sqrt{2}}\kappa s\psi_S^2 +\mbox{h.c.}  \non
&=& -\frac{1}{2} (\psi^0)^T Y_{\chi^0} \psi^0 + \mbox{h.c.} ,
\end{eqnarray}
where
\begin{eqnarray}
(\psi^0)^T=(-i\lambda^1,-i\lambda^2,\psi^1_{H_d},\psi^2_{H_u},\psi_S) .
\end{eqnarray}
The neutralinos are obtained by the unitary rotation $\psi^0_i=(Z_N)_{ij} \chi^0_j$,
where $Z_N$ diagonalizes the mass matrix $Y_{\chi^0}$.

Similar to the charged Higgs sector, the chargino sector of the NMSSM is the same
as in the MSSM with $\mu$ replaced by $\mu_{eff}$. The chargino masses are
obtained by the diagonalization of the mass matrix with
two unitary matrices $Z_-$ and $Z_+$:
\begin{eqnarray}
M_{\chi^C}=  \left(Z_-\right)^T\left(
\begin{array}{cc}
-M_2 &             \sqrt{2} m_W\sin\beta \\
\sqrt{2} m_W\cos\beta & -\mu_{eff}
\end{array}
\right) Z_+.
\end{eqnarray}

\section{Calculations of Wilson coefficients}
\label{Sec:CACQ}

In our calculations we consider the flavor mixing between  $\tilde
b$ and $\tilde s$, which make contributions to the dileptonic B
meson decays through gluino or neutralino loops. Following the
analysis in \cite{Hikasa87}, we  assume the flavors are diagonal
at tree level and the mixings are induced at loop level. Such
mixings can be parameterized by a small mixing parameter
$\epsilon_1$ which is dependent on some soft-breaking mass
parameters \cite{Hikasa87}. In our numerical calculations we input
$\epsilon_1=0.1$ for illustration. We perform the calculations in
the Feynman gauge and thus the Goldstone bosons will be involved
in the loop diagrams.

Since in the NMSSM the lighter mass eigenstate,
the CP-odd neutral Higgs boson $A_1$,
can be rather light, with a mass ranging from 100 MeV to the weak scale
\cite{reviewMSSM,hierarchy} ( using the package NMHDECAY \cite{nmhdecay},
we checked that such a light CP-odd Higgs boson $A_1$ is indeed allowed by
the LEP II data), in our calculations we pay special attention to this
wide mass range of $A_1$ and discriminate three cases.
\begin{itemize}

\item[(i)]  {\em Case A:} Heavy $A_1$.

For a heavy $A_1$, around weak scale, we integrate it out together with the other
heavy particles (Higgs bosons, top quark, $W^\pm$ and $Z$ bosons, sparticles)
at weak scale to obtain the Wilson coefficients.
The effective Hamiltonian describing $b\to sl^+l^-$ transition reads
\bea
{\cal H}_{eff}&=&-\frac{4G_F}{\sqrt{2}}V_{tb}V_{ts}^* \left[ \sum_{i=1}^{10}
    C_i(\mu_r){\cal O}_i(\mu_r)+\sum_{i=1}^{2} C_{Q_i}(\mu_r){\cal Q}_i(\mu_r)\right ]\ ,
\label{hamilton}
\eea
where ${\cal O}_i$ and ${\cal Q}_i$ are operators listed in \cite{Dai97,Xiong01},
and $C_i$ and $C_{Q_i}$ are respectively their Wilson coefficients,
and $\mu_r$ is the renormalization scale. Note that the most general Hamiltonian in
low-energy supersymmetry also contains the operators ${\cal O}_i^\prime$ and
${\cal Q}_i^\prime$ which respectively are the flipped chirality partners of
${\cal O}_i$ and ${\cal Q}_i$. However, they give negligible contributions
and thus are not considered in the final discussion of physical quantities \cite{Lunghi99}.

In this case, there are no new operators and we only need to calculate
the NMSSM contributions to these coefficients $C_{i}$ and $C_{Q_i}$ at the scale of $m_W$.
For the processes in our analysis, only $C_{7,9,10}$ and $C_{Q_{1,2}}$ are relevant.
The NMSSM contributions to $C_{7,9,10}$ are the same as in the MSSM, which are
computed in \cite{Bertolini91}.
For $C_{Q_{1,2}}$ the NMSSM contributions are different from the MSSM  contributions \cite{Xiong01}
and thus need to be calculated here. The Feynman diagrams we need to calculate are shown in
Fig.1, where the loops respectively involve the charged Higgs bosons,
charginos, gluinos and neutralinos.
From the calculations of these diagrams we obtain the Wilson coefficients
at $m_W$ scale, which are presented in the Appendix.

\begin{figure}[htb]
\scalebox{1.2}{\epsfig{file=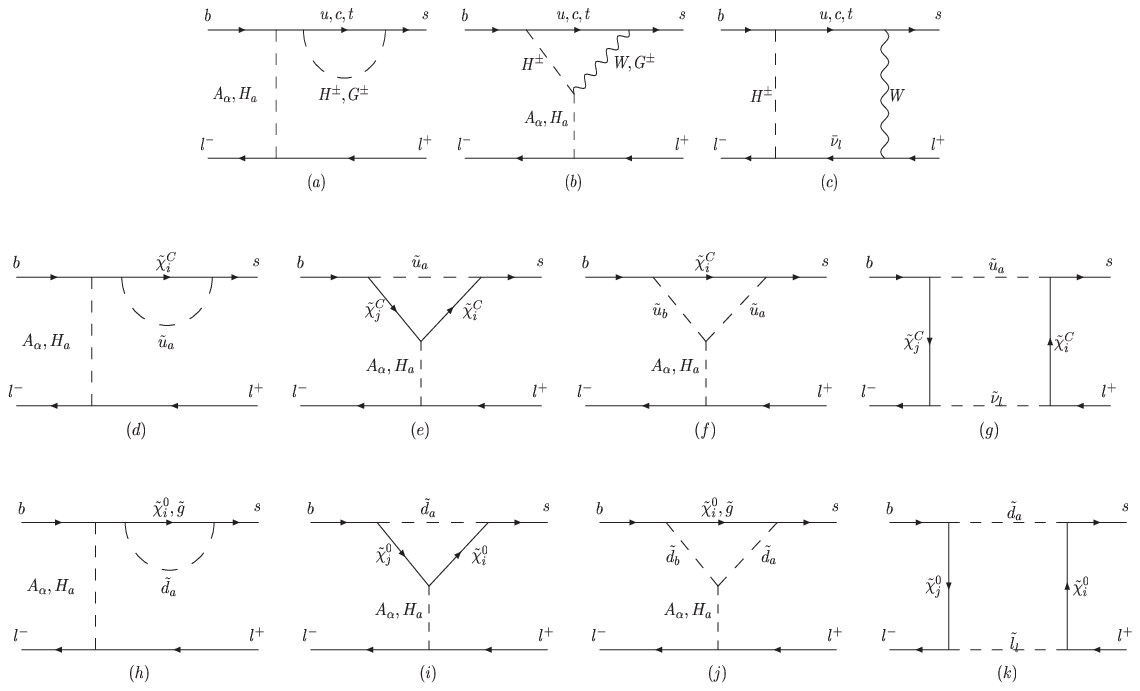}}
\vspace*{-0.6cm}
\caption{The Feynman diagrams which give the dominate contributions to $C_{Q_{1,2}}$:
(a-c) charged Higgs loops, (d-g) chargino loops, (h-k) gluino and neutralino loops.}
\label{fey-charged-higgs}
\end{figure}
For the calculation of the dileptonic B meson decays we need to
know the Wilson coefficients at the $m_{b}$ scale, which can be
obtained from the running of the coefficients at the $m_W$ scale
down to the $m_{b}$ scale. Such a running is governed by the
anomalous dimension which can be found in \cite{Xiong01}.

\item[(ii)] {\em Case B:} $A_1$ with an intermediate mass $m_b\ll m_{A_1}\ll m_W$.

At the $m_W$ scale we retain $A_1$ as an active field and thus we
have a new operator $\mathcal{O}_A$. After integrating out all
heavy particles at the $m_W$ scale we obtain the Wilson
coefficients including $C_A$ for $\mathcal{O}_A$. Then we work out
the anomalous dimensions and run the Wilson coefficients from the
$m_W$ scale to the $m_{A_1}$ scale. At the $m_{A_1}$ scale we
integrate out $A_1$, which makes an additional contribution to
$C_{Q_i}(m_{A_1})$. Finally, we run all the Wilson coefficients
from the  $m_{A_1}$ scale to the  $m_b$ scale.

The new operator $\mathcal{O}_A$ at the $m_W$ scale takes the form
\begin{eqnarray}
\mathcal{O}_A=i\frac{g_2}{16\pi^2}m_bm_W\bar{s}_L^\alpha b_R^\alpha A_1 .
\end{eqnarray}
From the calculations of the corresponding diagrams in Fig. 1 we obtain
the Wilson coefficient $C_A(m_W)$. It is composed of the charged Higgs loop
contribution, the chargino loop contribution, the neutralino loop contribution
and the gluino loop contribution, whose analytic expressions are given in the
Appendix.

For the running of the Wilson coefficients including $C_A$ from the $m_W$ scale
to the $m_{A_1}$ scale we work out the anomalous dimensions by calculating
one-loop diagrams with operator insertions. We find that all the Wilson
coefficients in Eq.(\ref{hamilton}) run in the same way as in the MSSM \cite{Xiong01},
while the new coefficient $C_{A}$ is not changed, i.e., $C_A(m_{A_1})=C_A(m_W)$.

When integrating out $A_1$ at the $m_{A_1}$ scale we find it gives a contribution
$\Delta C_{Q_2}(m_A)$ to the operator ${\cal Q}_2$
\bea
\Delta C_{Q_2}(m_{A_1})=-\frac{\delta_-}{2}\frac{v}{s}\frac{m_bm_l}{m^2_{A_1}}C_A(m_{A_1}).
\label{CAcoeff}
\eea
Finally, for the running of the Wilson coefficients  from the $m_{A_1}$ scale to the $m_b$ scale
the anomalous dimensions are the same as in the MSSM \cite{Xiong01}.

\item[(iii)] {\em Case C:} Super light $A_1$ with mass $m_{A_1}< m_b$.

In this case we retain $A_1$ as an active field in the entire
analysis. At the  $m_W$ scale we integrate out all heavy particles
and obtain the Wilson coefficients including $C_A$. Then we run
the coefficients down to the $m_b$ scale. At the $m_b$ scale the
effects of  $\mathcal{O}_A$ are represented by a change in
$C_{Q_2}$ \bea \Delta
C_{Q_2}(m_b)=\frac{\delta_-}{2}\frac{v}{s}\frac{m_bm_l}{p^2-m^2_{A_1}+im_{A_1}\Gamma_{A_1}}
C_A(m_b), \eea where $p$ is the momentum transfer and
$\Gamma_{A_1}$ is the total width of $A_1$. Since  $A_1$ can be
on-shell in this case, the effects of $A_1$ can be sizable even if
without $\tan\beta$ enhancement.
\end{itemize}
Note that the chargino loop contributions to $C_A$
were also calculated in \cite{hiller} where the
corresponding diagrams induced by the $A_1$-squark-squark
vertex are neglected since the author considered the
large $\tan\beta$ limit. In our numerical calculations
we used the full results by keeping all terms
and thus we also included the diagrams induced by the $A_1$-squark-squark
coupling although they contain no leading $\tan\beta$ terms.
Except for the case of a large $\tan\beta$, such diagrams induced by
the $A_1$-squark-squark coupling should be included since
the $A_1$-squark-squark coupling can arise from the F-term of the superpotential
and not suppressed by the singleness of $A_1$.
We checked that in the large $\tan\beta$ limit we can reproduce the
analytical result given in \cite{hiller} for the chargino-loop
contributions.

\section{Dileptonic B-meson decays in NMSSM}
\label{sec:decay}

With the effective Hamiltonian and the running of the Wilson coefficients
presented in the preceding section we calculate the inclusive decays
$B\to X_s \ell^+\ell^-$ and their forward-backward (FB) asymmetry,
as well as the exclusive decays $B_s\to \ell^+\ell^-\gamma$.
The formulas in terms of the Wilson coefficients can be found in
\cite{Dai97,Xiong01}.

Note that our supersymmetric contributions to the  Wilson coefficients are given
at one-loop level (next-to-leading order), while the SM  contributions are known
at two-loop level (next-to-next-to-leading order)\cite{Ghinculov04}.
In our numerical calculations we
consider the one-loop results for the NMSSM, while for the SM we also include
the two-loop results.

For the inclusive decays $B\to X_s \ell^+\ell^-$
we exclude the resonances $J/\Psi$ and $\Psi'$ contributions by using the same
cuts as in the experiments \cite{cut_ex}, i.e., the invariant dilepton mass
in the ranges
\begin{eqnarray} \label{cut}
(2m_l, 2.75{\rm ~GeV})\oplus (3.3{\rm ~GeV},3.39{\rm ~GeV})\oplus (3.84{\rm ~GeV},m_b),
\end{eqnarray}
so that our results can be compared with the experimental measurements.

For exclusive decays $B_s\to \ell^+\ell^-\gamma$ we follow
\cite{Aliev97,Xiong01} and consider the photon in
$B_s\to\ell^+\ell^-\gamma$ as a hard photon by imposing a cut on
the photon energy $E_\gamma$, which means that the radiated photon
can be detected in  the experiments. This cut requires
$E_\gamma\geq \delta~ m_{B_s}/2$ with $\delta = 0.02$. (Note that
for a soft photon both processes $B_s\to\ell^+\ell^-\gamma$ and
$B_s\to\ell^+\ell^-$ must be considered together and in this case
the infrared singular terms in $B_s\to \ell^+\ell^-\gamma$ are be
cancelled by the $O(\alpha_{em})$ virtual corrections in
$B_s\to\ell^+\ell^-$ \cite{Aliev97}.)

In our numerical calculations we perform a scan over the NMSSM parameter space
\begin{eqnarray}
&& 2\le \tan\beta \le 30, ~~-500 {\rm ~GeV} \le \mu_{eff} \le 500 {\rm ~GeV},~~\nonumber \\
&& -1 \le \lambda  \le 1, ~~ -1 \le  \kappa \le 1, ~~ \nonumber \\
&&  -50 {\rm ~GeV} \le  A_\lambda \le 50 {\rm ~GeV}, ~~
 -10 {\rm ~GeV} \le A_\kappa \le 10 {\rm ~GeV}
\end{eqnarray}
with fixed parameters for the sfermion and gaugino sector (500 GeV for all sfermions
and the gluino, and 200 GeV and 100 GeV for $SU(2)$ and $U(1)$ gaugino masses $M_2$
and $M_1$, respectively).
In our scan we consider the following constraints:
\begin{itemize}
\item[(1)] The LEP2 constraints by using the package NMHDECAY \cite{nmhdecay}.
\item[(2)] The  constraints from $B\to X_s\gamma$ which stringently constrain
           the effective coefficient $C_7^{eff}$.
       For the experimental result we use the world average value \cite{HFAG}
    \begin{equation}
    Br(B \to X_s\gamma)|_{exp}=(3.55\pm0.24^{+0.09}_{-0.10}\pm 0.03)\times 10^{-4}.
    \label{bsrex}
    \end{equation}
\item[(3)] The constraints from $B_s\to\mu^+\mu^-$, which
constrain the
           Wilson coefficient $C_{A}$ of the light pseudoscalar operator \cite{hiller,Xiong01}.
           The experimental result is given by \cite{PDG}
    \begin{eqnarray}
    Br(B_s\to\mu^+\mu^-)<1.5\times 10^{-7}~~(90\% ~C.L.).
    \end{eqnarray}
\end{itemize}

Among the relevant dileptonic decays
experiment data is only available for $Br(B\to X_s \mu^+ \mu^-)$,
which is given by \cite{PDG}
\begin{eqnarray}
Br(B\to X_s \mu^+ \mu^-)=(4.3\pm 1.2) \times 10^{-6}.
\end{eqnarray}
In displaying our numerical results we will show this bound and use it to constrain
the parameter space.
For other dileptonic decay branching ratios, with no experiment data available,
we will compare the NMSSM predictions with the SM values given by
\begin{eqnarray}
&& Br(B\to X_s \tau^+ \tau^-)=4.43\times 10^{-8}, \\
&& Br(B_s\to \gamma \mu^+ \mu^-)=1.33\times 10^{-8}, \\
&& Br(B_s\to \gamma \tau^+ \tau^-)=1.35\times 10^{-8}.
\end{eqnarray}
Note that the SM prediction for $Br(B\to X_s \tau^+ \tau^-)$ was also given
in \cite{Ali1996}. But our result is different from theirs because
it is very sensitive to the cuts around the resonances $J/\psi$ and $\psi'$.
While our cuts are chosen as in
Eq.(\ref{cut}), we cannot find the corresponding cuts used in \cite{Ali1996}.
We can easily reproduce the result in \cite{Ali1996} by varying
the cuts.

\begin{figure}[htb]
\scalebox{1.1}{\epsfig{file=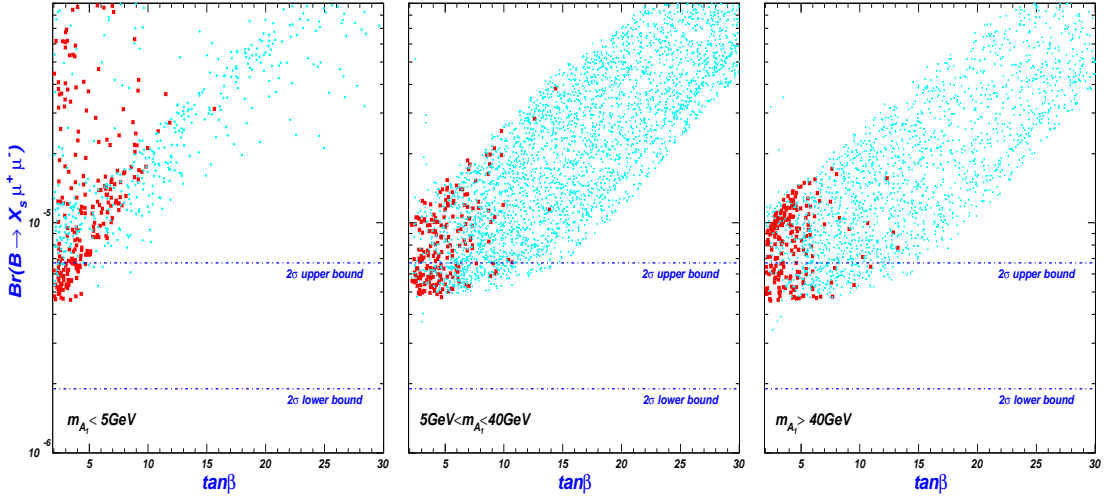}}
\vspace*{-0.5cm}
\caption{Scatter plots for the branching ratio of $B\to X_s\mu^+\mu^-$ versus $\tan\beta$:
the left panel is for a super light $A_1$ ($m_{A_1}<$ 5GeV), the middle  panel  is
for an intermediately heavy $A_1$ (5 GeV $<m_{A_1}<$ 40 GeV) and
the right  panel  is for a heavy $A_1$ at $m_W$ scale.
The dark (red) points are allowed by $b\to s \gamma $, while
the light (sky-blue) points are excluded by  $b\to s \gamma $. }
\label{bsmm_tanb}
\end{figure}

In Fig.\ref{bsmm_tanb} we show the scatter plots of the
branching ratio for $B\to X_s\mu^+\mu^-$ versus $\tan\beta$.
Here we present the results for the three cases:
a super light $A_1$ ($m_{A_1}<$5GeV), an intermediately heavy $A_1$
(5 GeV$<m_{A_1}<$ 40 GeV) and a heavy  $A_1$ at $m_W$ scale.
In order to see how stringent the $b\to s \gamma $ constraints are
we display the scatter plots with and without the $b\to s \gamma $ constraints.
From this figure we make the following observations:
(1) The branching ratio can be greatly enhanced by large $\tan\beta$.
(2) $b\to s\gamma$ constraints are quite stringent and can exclude a
    large part of the parameter space, typically with large  $\tan\beta$.
(3) In the parameter space allowed by $b\to s\gamma$ the decay
    $B\to X_s\mu^+\mu^-$ can still be greatly enhanced, especially in
    case of a super light $A_1$. The $2\sigma$ experimental
    bound on $B\to X_s\mu^+\mu^-$ can further exclude a large part of the parameter
    space.  Almost no points with $\tan\beta>15$ in the parameter space survive all
    the constraints.
From the left panel of Fig.\ref{bsmm_tanb} we see that some part of the
parameter space with a super light $A_1$
is still allowed by $b\to s\gamma$ and $B\to X_s\mu^+\mu^-$.

Let us take a look on the constraints from the process $B_s\to \mu^+\mu^-$,
whose branching ratio is given by \cite{xbsmm}
\begin{eqnarray}
Br(B_s\to \mu^+\mu^-)&=&1.2\times 10^{-7}\left[ \frac{\tau_{B_s}}{1.49ps}\right]
\left[\frac{f_{B_s}}{245MeV}\right]^2\left| \frac{V_{ts}}{0.04}\right|^2
\left[\frac{m_{B_s}}{5.37GeV}\right]^3 \nonumber \\
&& \times \left[C^2_{Q_1}+\left(C_{Q_2}+2\frac{m_\mu}{m_{B_s}}C_{10}\right)^2\right].
 \label{eq28}
\end{eqnarray}
We see that the contributions are from $C_{10}$ and $C_{Q_{1,2}}$. While the
contribution from  $C_{10}$ is suppressed by the factor $m_\mu/m_{B_s}$,
the contributions from $C_{Q_{1,2}}$ can be enhanced by large $\tan\beta$
($C_{Q_{1,2}}$ contain terms which are proportional to $\tan^3\beta$, as shown in the Appendix).
This feature can be seen from Fig.\ref{b2mm} in which we set aside the $b\to s\gamma$
constraints and illustrated the $B_s\to \mu^+\mu^-$ constraints.
We see that, similar to the $b\to s\gamma$ constraints,
the $B_s\to \mu^+\mu^-$ constraints are stringent for a large $\tan\beta$.
If we impose the $b\to s\gamma$  constraints which exclude a very
large $\tan\beta$, then the further  constraints from $B_s\to \mu^+\mu^-$
are stringent only for the parameter space with a very light $A_1$.

\begin{figure}[htb]
\scalebox{1.1}{\epsfig{file=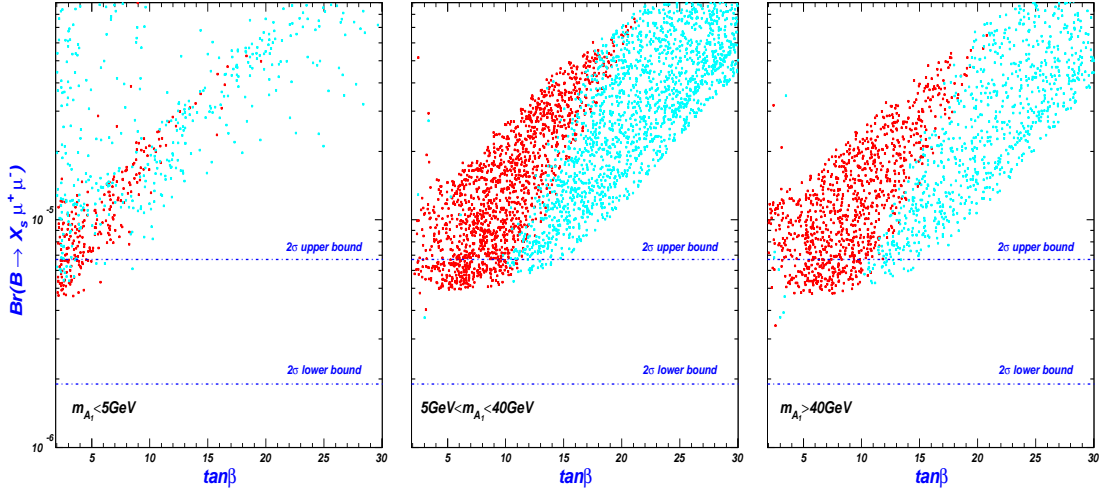}}
\vspace*{-0.5cm}
\caption{Same as Fig.\ref{bsmm_tanb}, but with the $B_s\to \mu^+\mu^-$ constraints.
The dark (red) points are allowed by $B_s\to \mu^+\mu^- $, while
the light (sky-blue) points are excluded by  $B_s\to \mu^+\mu^-$. }
\label{b2mm}
\end{figure}

\begin{figure}[htb]
\scalebox{0.9}{\epsfig{file=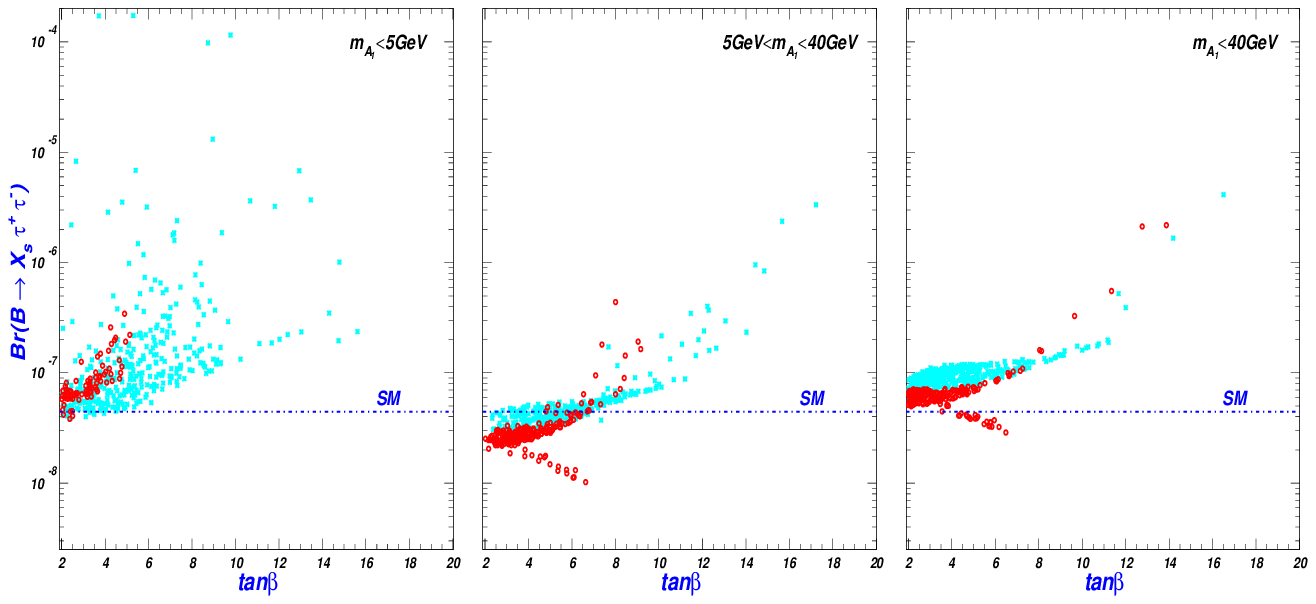}}
\scalebox{0.9}{\epsfig{file=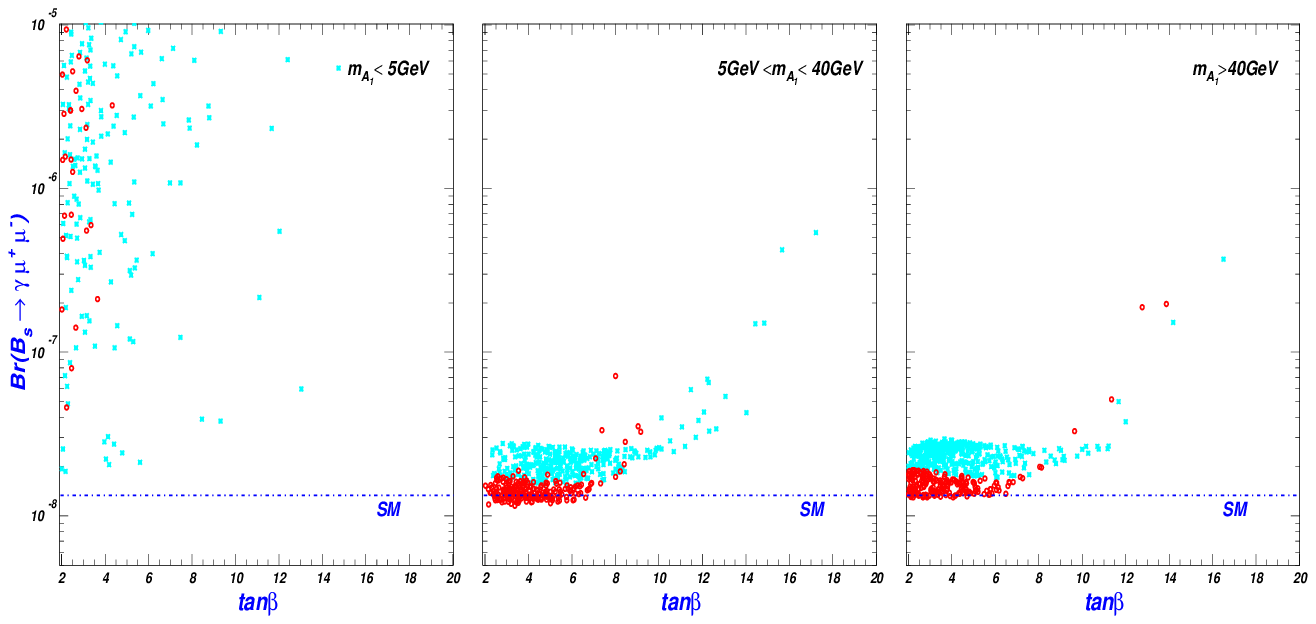}}
\scalebox{0.9}{\epsfig{file=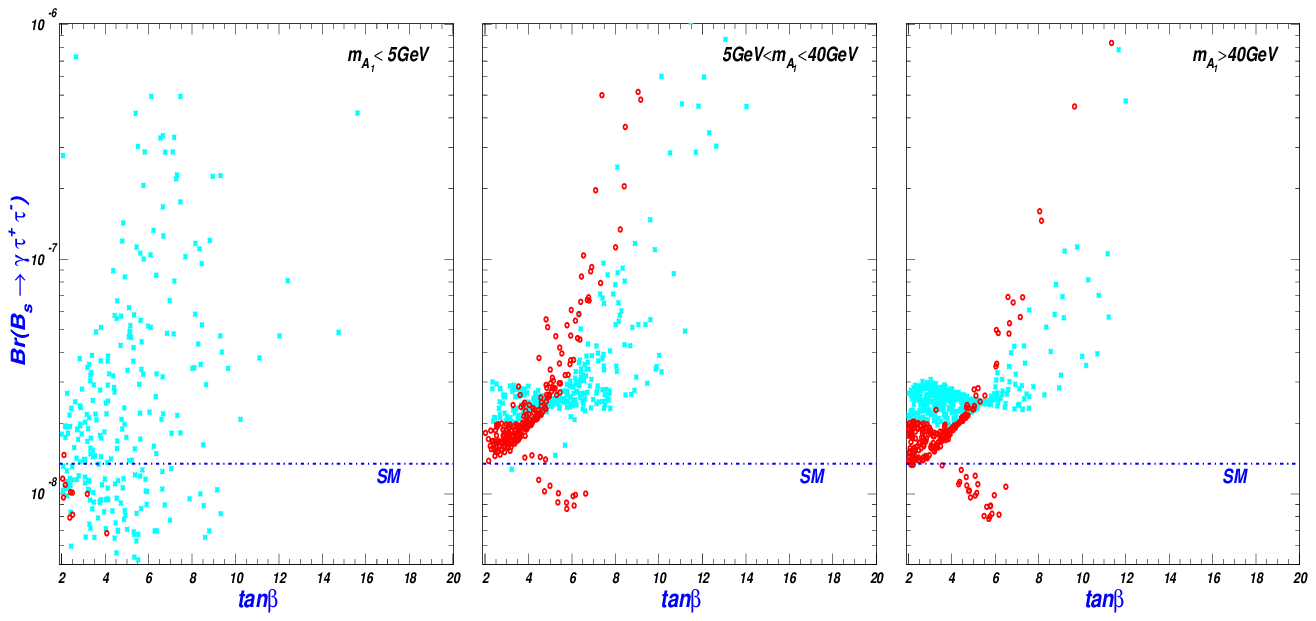}}
\vspace*{-0.4cm}
\caption{Same as Fig.\ref{bsmm_tanb}, but for $B\to X_s \tau^+\tau^-$,
 $B_s\to \gamma \mu^+\mu^-$ and  $B_s\to \gamma \tau^+\tau^-$.
The dark (red) points are allowed by $B\to X_s\mu^+\mu^-$,
while the light (sky-blue) points are excluded by $B\to X_s\mu^+\mu^-$.}
\label{bstt_tanb}
\end{figure}
The results for other dileptonic decay branching ratios, for which
no experiment data are available, are presented in
Fig.\ref{bstt_tanb}. To see how stringent the constraints from
$B\to X_s\mu^+\mu^-$ are we display the scatter plots with and
without such constraints (all the points satisfy $b\to s\gamma$
and $B_s\to \mu^+\mu^-$).

From Fig.\ref{bstt_tanb} we see that under the constraint from
$B\to X_s\mu^+\mu^-$, the branching ratio of $B\to X_s \tau^+ \tau^-$
does not deviate
significantly from the SM value. The reason is that these two
decays are highly correlated except for the contributions of
$C_{Q_{1,2}}$ and $C_A$ which are dependent on the lepton mass. If
the contributions of $C_{Q_{1,2}}$ and $C_A$ are dominant, then
$Br(B\to X_s \tau^+ \tau^-)$ should not be severely constrained by
$B\to X_s\mu^+\mu^-$. As discussed below Eq.(\ref{eq28}), the
contributions of $C_{Q_{1,2}}$ and $C_A$ are important only for
very large $\tan\beta$ which is not allowed by $b\to s\gamma$. As
a result, the contributions of $C_{Q_{1,2}}$ and $C_A$ are not
dominant and thus $B\to X_s \tau^+ \tau^-$ is highly correlated to
$B\to X_s\mu^+\mu^-$.

Note that for $B\to X_s \tau^+ \tau^-$ it may be rather
challenging to disentangle the NMSSM effects from the SM value in
future experiments. One reason is that, as discussed above, the
NMSSM effects are no longer so sizable under the constraint of
$B\to X_s\mu^+\mu^-$. The other reason is that the SM prediction
has its own uncertainty. If we consider the uncertainty of the
input SM parameters, we can obtain the uncertainty (about $20\%$
as found in \cite{Ali1996}) of the SM prediction. But in
Fig.\ref{bstt_tanb} we did not show such an uncertainty of the SM
value because for all the results, both NMSSM and SM, we used a
same set of the SM parameters without allowing them to vary in the
uncertainty range. Since the SM parameters are involved in both
the NMSSM and SM values, all the results are subject to some
uncertainty if we consider the uncertainty of the SM parameters.
Of course, such  uncertainties will deteriorate the observability
of the NMSSM effects.

\begin{figure}[htb]
\scalebox{0.4}{\epsfig{file=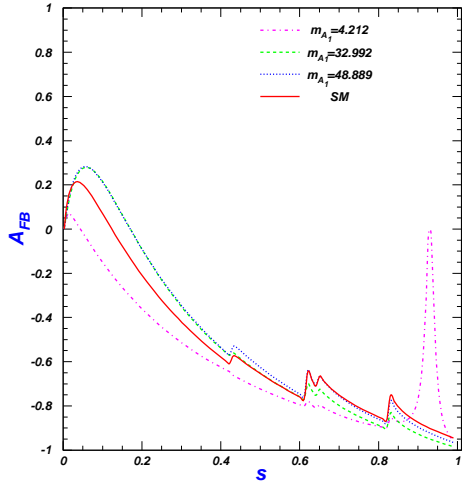}} \caption{The
forward-backward asymmetry of $B\to X_s \mu^+\mu^-$ versus
 $s=p^2/m^2_b$
($p^2$ is the invariant mass of $\mu^+$ and $\mu^-$).}
\label{A_fb}
\end{figure}
Finally, in  Fig.\ref{A_fb} we show the results for the forward-backward asymmetry
in $B\to X_s \mu^+\mu^-$ under the constraints from $B\to X_s\mu^+\mu^-$.

\section{Summary}
\label{sec:summary}
In the framework of the NMSSM we examined  the rare dileptonic
decays $B\to X_s\ell^+\ell^-$ and $B_s\to\ell^+\ell^-\gamma$
paying particular attention to the light CP-odd Higgs boson.
We found that in the parameter space allowed by current experiments, such as LEP II and
$b\to s \gamma$, the branching ratios of these rare decays can be greatly enhanced and
thus the experimental data on $B\to X_s \mu^+ \mu^-$  further stringently
constrains the parameter space, especially with a super light CP-odd Higgs
boson and large $\tan\beta$.
In the surviving parameter space we gave the NMSSM predictions for other
unmeasured dileptonic decays which may hopefully be measured at the future LHCb or
super B factory.

\section*{Acknowledgment}
We thank J. J. Cao, C. S. Huang, K. Hikasa and  K. Tobe for discussions.
This work was supported in part by National Natural
Science Foundation of China (NNSFC) under grant No. 10725526 and 10635030,
by the US Department of Energy, Division of High Energy Physics
under No. DE-FG02-91-ER4086 and by the JSPS invitation program 
under No. S-08034. 

\appendix

\section{Wilson Coefficients}
The Wilson coefficients $C_7$, $C_9$ and $C_{10}$ in the NMSSM are
the same as in the MSSM \cite{Bertolini91}. Here we present the
new coefficient $C_A$, and $C_{Q_1}$ and $C_{Q_2}$, whose
predictions in the NMSSM are different from the MSSM. We checked
that we can analytically reproduce the MSSM results
\cite{Bertolini91,Xiong01,other} (However, the NMSSM results can
not explicitly reduce to the MSSM results  by simply dropping out
the singlet $\hat S$ (say setting $\lambda=\kappa=0$) because the
$\mu$-term is generated by $\hat S$).

Although in our numerical calculations we used the complete results by
keeping all terms, here, for simplicity, we only present the terms which can be
enhanced by large $\tan\beta$.
At the $m_W$ scale each Wilson coefficient is composed of the charged Higgs loop
contribution from Fig.1(a-c),
the chargino loop contribution from Fig.1(d-g),
the neutralino and gluino loop contribution from  Fig.1(h-k).
\vspace*{0.5cm}

For the charged Higgs contributions:
\small
\begin{eqnarray}
&& C^{H^\pm}_{A}=-\frac{i\lambda A_\lambda}{g_2m_W}\tan\beta F_1(x_{H^{\pm}t},x_{Wt})\\
&& C_{Q_{1}}^{H^{\pm}}=-\frac{m_{b}m_{l}}{4m_{H_{a}}^{2}}\tan^{2}\beta
   \biggl[\frac{m^{2}_{H^{\pm}}}{m^{2}_{W}}U^{H}_{a1}U^{H}_{a1}F_{1}(x_{H^{\pm}t},x_{Wt})
   +\frac{m^{2}_{t}m_{H_{a}}^{2}}{m^{2}_{W}m^{2}_{H^{\pm}}}F_{1}(x_{tH^{\pm}},x_{tW})\biggl]\\
&& C_{Q_{2}}^{H^{\pm}}=\frac{m_{b}m_{l}}{4m_{A_{\alpha}}^{2}}\tan^{2}\beta
   \biggl[\biggl(\frac{m^{2}_{H^{\pm}}}{m^{2}_{W}}U^{A}_{\alpha1}U^{A}_{\alpha1}
   +\delta_{\alpha 2}U^{A}_{\alpha 1}\biggl) F_{1}(x_{H^{\pm}t},x_{Wt}) \nonumber\\
&& \hspace*{4cm}
  +\frac{m^{2}_{t}m_{A_{\alpha}}^{2}}{m^{2}_{W}m^{2}_{H^{\pm}}}F_{1}(x_{tH^{\pm}},x_{tW})\biggl]
\end{eqnarray}
\normalsize
Note that although $H^\pm W^\pm A_1$ vertex has $1/\tan\beta$ suppression
by the singleness of $A_1$, $H^\pm G^\pm A_1$ vertex which comes from the
soft term $A_\lambda \lambda SH_u^1H_d^2$ does not have such $1/\tan\beta$ suppression.
Thus the contribution to $C_A$ from the loop involving $H^\pm$ and $G^\pm$ with
$H^\pm G^\pm A_1$ coupling (in the Feynman gauge) is proportional to $\tan\beta$.

For the chargino contributions:
\small
\begin{eqnarray}
C^{\chi^C}_A&=&i\frac{\tan\beta}{\sqrt{2}}\Gamma_{1}(i,i,j,l)
   \Biggl\{\delta_{-}\delta_{lj}\frac{v}{s}x^{1/2}_{\chi_{j}^{C}W}
    P_{1}\left(x_{\tilde{t}_{i-1}\tilde{\chi}_{j}^{C}}\right)\non
&&    -\biggl[R_{1jl}x^{1/2}_{\chi_{j}^{C}\chi_{l}^{C}}F_{1}
   \left(x_{\tilde{t}_{i-1}\tilde{\chi}_{l}^{C}}, x_{\tilde{\chi}_{j}^{C}\tilde{\chi}_{l}^{C}}\right)
   -R^*_{1lj}F_{2}
   \left(x_{\tilde{t}_{i-1}\tilde{\chi}_{l}^{C}}, x_{\tilde{\chi}_{j}^{C}\tilde{\chi}_{l}^{C}}\right)
   \biggl]\Biggl\}, \\
C_{Q_{1}}^{\widetilde{\chi}^{\pm}}&=&\frac{m_{b}m_{l}}{4m_{H_{a}}^{2}}\tan^{2}\beta
        \sum_{i,k=1}^{3}\sum_{j,l=1}^{2}\Gamma_{1}(i,k,j,l)
        \Biggl\{\frac{\sqrt{2}U^{H}_{a1}U^{H}_{a1}m_{\chi_{j}^{C}}}{m_{W}\cos\beta}\delta_{ik}\delta_{lj}
      P_{1}(x_{\tilde{t}_{i-1}\tilde{\chi}_{j}^{C}})\nonumber \\
&& -\frac{2\sqrt{2}U^{H}_{a1}}{g_{2}}\delta_{ik}\biggl[Q_{alj}^{*}
   F_{2}(x_{\tilde{t}_{i-1}\tilde{\chi}_{l}^{C}},x_{\tilde{\chi}_{j}^{C}\tilde{\chi}_{l}^{C}})
   +\frac{m_{\chi_{j}^{C}}}{m_{\chi_{l}^{C}}}Q_{ajl}
    F_{1}(x_{\tilde{t}_{i-1}\tilde{\chi}_{l}^{C}},
    x_{\tilde{\chi}_{j}^{C}\tilde{\chi}_{l}^{C}})\biggl]\nonumber \\
&& +\frac{2\sqrt{2}U^{H}_{a1}T^{aik}_{2}m_{\chi_{j}^{C}}}{m^{2}_{\widetilde{t}_{k-1}}}\delta_{lj}
    F_{1}(x_{\tilde{t}_{i-1}\tilde{t}_{k-1}},x_{\tilde{\chi}_{j}^{C}\tilde{t}_{k-1}})\nonumber \\
&&  +\frac{m^{2}_{H_{a}}}{m_{\chi_{j}^{C}}^{2}}\delta_{ik}\Biggl[Z_{-}^{2j}Z_{+}^{1l}
      F_{4}(x_{\tilde{t}_{i-1}\tilde{\chi}_{j}^{C}},x_{\tilde{\chi}_{l}^{C}\tilde{\chi}_{j}^{C}},
      x_{\tilde{\nu}\tilde{\chi}_{l}^{C}}) \nonumber\\
&&
      -\frac{m_{\chi_{l}^{C}}}{m_{\chi_{j}^{C}}}Z_{-}^{2l^{*}}Z_{+}^{1j^{*}}
      F_{3}(x_{\tilde{t}_{i-1}\tilde{\chi}_{j}^{C}},x_{\tilde{\chi}_{l}^{C}\tilde{\chi}_{j}^{C}},
      x_{\tilde{\nu}\tilde{\chi}_{l}^{C}})\Biggl]\Biggl\}, \label{A5}
\end{eqnarray}
\begin{eqnarray}
C_{Q_{2}}^{\widetilde{\chi}^{\pm}}&=&-\frac{m_{b}m_{l}}{4m_{A_{\alpha}}^{2}}\tan^{2}
    \beta\sum_{i,k=1}^{3}\sum_{j,l=1}^{2}\Gamma_{1}(i,k,j,l)
     \Biggl\{\frac{\sqrt{2}U^{A}_{\alpha1}U^{A}_{\alpha1}m_{\chi_{j}^{C}}}{m_{W}\cos\beta}
      \delta_{ik}\delta_{lj} P_{1}(x_{\tilde{t}_{i-1}\tilde{\chi}_{j}^{C}})\nonumber \\
&& -\frac{2\sqrt{2}U^{A}_{\alpha1}}{g_{2}}\delta_{ik}\biggl[-R_{alj}^{*}
   F_{2}(x_{\tilde{t}_{i-1}\tilde{\chi}_{l}^{C}},x_{\tilde{\chi}_{j}^{C}\tilde{\chi}_{l}^{C}})
   +\frac{m_{\chi_{j}^{C}}}{m_{\chi_{l}^{C}}}R_{ajl}
   F_{1}(x_{\tilde{t}_{i-1}\tilde{\chi}_{l}^{C}},
   x_{\tilde{\chi}_{j}^{C}\tilde{\chi}_{l}^{C}})\biggl]\nonumber \\
&& -\frac{\sqrt{2}U^{A}_{\alpha1}T^{\alpha ik}_{1}m_{t}m_{\chi_{j}^{C}}}{m_{W}
   m^{2}_{\widetilde{t}_{k-1}}}\delta_{lj}
   F_{1}(x_{\tilde{t}_{i-1}\tilde{t}_{k-1}},x_{\tilde{\chi}_{j}^{C}\tilde{t}_{k-1}})\nonumber\\
&& +\frac{m^{2}_{A_{\alpha}}}{m_{\chi_{j}^{C}}^{2}}\delta_{ik}\Biggl[Z_{-}^{2j}Z_{+}^{1l}
   F_{4}(x_{\tilde{t}_{i-1}\tilde{\chi}_{j}^{C}},x_{\tilde{\chi}_{l}^{C}\tilde{\chi}_{j}^{C}},
   x_{\tilde{\nu}\tilde{\chi}_{l}^{C}}) \nonumber\\
&&
   -\frac{m_{\chi_{l}^{C}}}{m_{\chi_{j}^{C}}}Z_{-}^{2l^{*}}Z_{+}^{1j^{*}}
   F_{3}(x_{\tilde{t}_{i-1}\tilde{\chi}_{j}^{C}},x_{\tilde{\chi}_{l}^{C}\tilde{\chi}_{j}^{C}},
   x_{\tilde{\nu}\tilde{\chi}_{l}^{C}})\Biggl]\Biggl\} \label{A6}
\end{eqnarray}
\normalsize
Note that here $C_{Q_{1,2}}$ contain terms which can be enhanced by $\tan^3\beta$ (the overall
factor  $\tan^2\beta$ multiplied by $1/\cos\beta$ gives $\tan^3\beta$ in large $\tan\beta$ limit).

For neutralino contributions:
\small
\begin{eqnarray}
C^{\chi^0}_{A}&=&-\frac{i}{V_{tb}V_{ts}^{*}}\frac{\tan\beta}{\cos\theta_{W}}
   N'_{j}\biggl\{Z_{N}^{3k^{*}}T_{D}^{i2}T_{D}^{i1^{*}}
 \biggl[R_{1jk}^{R^{''}}T_{D}^{i2} x^{1/2}_{\chi_{k}^{0}\chi_{j}^{0}}
   F_{1}(x_{\tilde{b}_{i-1}\tilde{\chi}_{j}^{0}},x_{\tilde{\chi}_{k}^{0}\tilde{\chi}_{j}^{0}})
   \non
&&    -R_{1jk}^{R^{*''}}F_{2}(x_{\tilde{b}_{i-1}\tilde{\chi}_{j}^{0}},
   x_{\tilde{\chi}_{k}^{0}\tilde{\chi}_{j}^{0}})\biggl]
 +T_{3}^{1ik'}\Gamma_{2}(i,j,k') (x_{\chi_{j}^{0}
   \widetilde{b}_{k'-1}}x_{W\widetilde{b}_{k'-1}})^{1/2}
 \non
&&\times
  F_{1}\left(x_{\tilde{b}_{i-1}\tilde{b}_{k'-1}},x_{\chi_{j}^{0}\tilde{b}_{k'-1}}\right)
 +\delta_{ik'}\frac{\delta_{-}\cot\beta}{\sqrt{2}}\frac{v}{s}x^{1/2}_{\chi_{j}^{0}b}
   \Gamma_{2}(i,j,k') P_{1}\left(x_{\tilde{b}_{i-1}
   \tilde{\chi}_{j}^{0}}\right) \biggl\}
\end{eqnarray}
\begin{eqnarray}
C_{Q_{1}}^{\widetilde{\chi}^{0}}&=&-\frac{1}{K_{tb}K_{ts}^{*}}\frac{m_{b}m_{l}}{4m_{H_{a}}^{2}
      \cos\theta_{w}}\tan^{2}\beta
      \sum_{i,k'=1}^{3}\sum_{j,k=1}^{5}\sum_{m=1}^{6}N'_{j}
      \biggl\{\frac{\sqrt{2}U^{H}_{a1}U^{H}_{a1}m_{\chi_{j}^{0}}}{m_{b}}\delta_{ik'}
         \Gamma_{2}(i,j,k')
    P_{1}(x_{\tilde{b}_{i-1}\tilde{\chi}_{j}^{0}})\nonumber \\
&&  -\frac{2U^{H}_{a1}Z_{N}^{3k^{*}}}{g_2}\biggl[Q_{ajk}^{L''}T_{D}^{i2}T_{D}^{i1^{*}}
    F_{2}(x_{\tilde{b}_{i-1}\tilde{\chi}_{j}^{0}},x_{\tilde{\chi}_{k}^{0}\tilde{\chi}_{j}^{0}})
    \nonumber\\
&&
    +Q_{ajk}^{R''}(\frac{m_{\chi_{k}^{0}}}{m_{\chi_{j}^{0}}}T_{D}^{i2}T_{D}^{i1^{*}}
    +\frac{m_b}{m_{\chi_{j}^{0}}}T_{D}^{i3}T_{D}^{i1^{*}})
    F_{1}(x_{\tilde{b}_{i-1}\tilde{\chi}_{j}^{0}},x_{\tilde{\chi}_{k}^{0}
    \tilde{\chi}_{j}^{0}})\biggl]\nonumber \\
&&  +2U^{H}_{a1}\frac{m_{\chi_{j}^{0}}}{m^{2}_{\widetilde{b}_{k'-1}}}
    (\frac{1}{\sqrt{2}}T_{4}^{aik'}\Gamma_{2}(i,j,k')
    +T_{5}^{aik'}Z_{N}^{3j^{*}}T_{D}^{i1^{*}}T_{D}^{k'2})
    F_{1}(x_{\tilde{b}_{i-1}\tilde{b}_{k'-1}},x_{\chi_{j}^{0}\tilde{b}_{k'-1}})\nonumber \\
&& +\frac{m^{2}_{H_{a}}}{m_{\chi_{k}^{0}}^{2}}
   \biggl[\biggl(\frac{m_{\chi_{j}^{0}}}{m_{\chi_{k}^{0}}}
   \Gamma_{6}(i,k)Z_{L}^{(I+3)m^{*}}Z_{L}^{Im}Z_{N}^{3k^{*}}Z_{N}^{3j^{*}}-\Gamma_{7}(i,k,j)\biggl)
   F_{3}(x_{\tilde{b}_{i-1}\tilde{\chi}_{j}^{0}},x_{\tilde{\chi}_{k}^{0}\tilde{\chi}_{j}^{0}},
   x_{\tilde{l}_{m}\tilde{\chi}_{j}^{0}})\nonumber \\
&& -\biggl(\Gamma_{6}(i,k)Z_{L}^{(I+3)m}Z_{L}^{Im^{*}}Z_{N}^{3k}Z_{N}^{3j}-\Gamma_{7}(i,j,k)\biggl)
   F_{4}(x_{\tilde{b}_{i-1}\tilde{\chi}_{j}^{0}},x_{\tilde{\chi}_{k}^{0}\tilde{\chi}_{j}^{0}},
   x_{\tilde{l}_{m}\tilde{\chi}_{j}^{0}})\biggl]\biggl\},\\
C_{Q_{2}}^{\widetilde{\chi}^{0}}&=&\frac{1}{K_{tb}K_{ts}^{*}}
    \frac{m_{b}m_{l}}{4m_{A_{\alpha}}^{2}\cos\theta_{w}}\tan^{2}\beta
    \sum_{i,k'=1}^{3}\sum_{j,k=1}^{5}\sum_{m=1}^{6}N'_{j}
    \biggl\{\frac{\sqrt{2}U^{A}_{\alpha1}U^{A}_{\alpha1}m_{\chi_{j}^{0}}}{m_{b}}\delta_{ik'}
   \Gamma_{2}(i,j,k') P_{1}(x_{\tilde{b}_{i-1}\tilde{\chi}_{j}^{0}})\nonumber \\
&& +\frac{2U^{A}_{\alpha1}Z_{N}^{3k^{*}}}{g_2}\biggl[R_{\alpha jk}^{L''}T_{D}^{i2}T_{D}^{i1^{*}}
   F_{2}(x_{\tilde{b}_{i-1}\tilde{\chi}_{j}^{0}},x_{\tilde{\chi}_{k}^{0}
   \tilde{\chi}_{j}^{0}}) \nonumber\\
&&
 +R_{\alpha jk}^{R''}(\frac{m_{\chi_{k}^{0}}}{m_{\chi_{j}^{0}}}T_{D}^{i2}T_{D}^{i1^{*}}
   +\frac{m_b}{m_{\chi_{j}^{0}}}T_{D}^{i3}T_{D}^{i1^{*}})
   F_{1}(x_{\tilde{b}_{i-1}\tilde{\chi}_{j}^{0}},
   x_{\tilde{\chi}_{k}^{0}\tilde{\chi}_{j}^{0}})\biggl]\nonumber\\
&& +2U^{A}_{\alpha1}\frac{m_{\chi_{j}^{0}}}{m^{2}_{\widetilde{b}_{k'-1}}}
   T_{3}^{\alpha ik'}\Gamma_{2}(i,j,k')
   F_{1}(x_{\tilde{b}_{i-1}\tilde{b}_{k'-1}},x_{\chi_{j}^{0}\tilde{b}_{k'-1}})\nonumber \\
&& -\frac{m^{2}_{A_{\alpha}}}{m_{\chi_{k}^{0}}^{2}}
   \biggl[\biggl(\frac{m_{\chi_{j}^{0}}}{m_{\chi_{k}^{0}}}
   \Gamma_{6}(i,k)Z_{L}^{(I+3)m^{*}}Z_{L}^{Im}Z_{N}^{3k^{*}}Z_{N}^{3j^{*}}-\Gamma_{7}(i,k,j)\biggl)
   F_{3}(x_{\tilde{b}_{i-1}\tilde{\chi}_{j}^{0}},x_{\tilde{\chi}_{k}^{0}\tilde{\chi}_{j}^{0}},
   x_{\tilde{l}_{m}\tilde{\chi}_{j}^{0}})\nonumber\\
&& -\biggl(\Gamma_{6}(i,k)Z_{L}^{(I+3)m}Z_{L}^{Im^{*}}Z_{N}^{3k}Z_{N}^{3j}-\Gamma_{7}(i,j,k)\biggl)
   F_{4}(x_{\tilde{b}_{i-1}\tilde{\chi}_{j}^{0}},x_{\tilde{\chi}_{k}^{0}\tilde{\chi}_{j}^{0}},
   x_{\tilde{l}_{m}\tilde{\chi}_{j}^{0}})\biggl]\biggl\}
\end{eqnarray}
\normalsize
For gluino contributions:
\small
\begin{eqnarray}
C^{\tilde{g}}_{A}&=&\frac{i}{V_{tb}V_{ts}^{*}}\frac{8g_{3}^{2}}{3g^{2}_{2}}\tan\beta
  T_{D}^{j1^{*}} \biggl[\delta_{ij}\frac{v}{s}
  \frac{\delta_{-}}{\tan\beta}x^{1/2}_{\tilde{g}b}
  P_{1}(x_{\tilde{b}_{i-1}\tilde{g}}) \nonumber\\
&& \hspace*{4cm}
  +T_{3}^{1ji}(x_{\tilde{g}\tilde{b}_{j-1}}x_{W{\tilde{b}_{j-1}}})^{1/2}
   F_{1}(x_{\tilde{b}_{i-1}\tilde{b}_{j-1}},x_{\tilde{g}\tilde{b}_{j-1}})\biggl], \\
C_{Q_{1}}^{\widetilde{g}}&=&\frac{1}{K_{tb}K_{ts}^{*}}\frac{m_{b}m_{l}}{4}\frac{16g_{3}^{2}
   m_{\widetilde{g}}}{3g^{2}_{2}m_{H_{a}}^{2}}\tan^{2}\beta
   \sum_{i,j=1}^{3}T_{D}^{i3}T_{D}^{j1^{*}}
  \Biggl[\frac{U^{H}_{a1}U^{H}_{a1}}{m_{b}}\delta_{ij}
   P_{1}(x_{\tilde{b}_{i-1}\tilde{g}}) \nonumber\\
&& \hspace*{6cm}
   +\frac{U^{H}_{a1}T_{4}^{aji}}{m_{\widetilde{b}_{j-1}}^{2}}
   F_{1}(x_{\tilde{b}_{i-1}\tilde{b}_{j-1}},x_{\tilde{g}\tilde{b}_{j-1}})\Biggl] ,\\
C_{Q_{2}}^{\widetilde{g}}&=&-\frac{1}{K_{tb}K_{ts}^{*}}\frac{m_{b}m_{l}}{4}
   \frac{16g_{3}^{2}m_{\widetilde{g}}}{3g^{2}_{2}m_{A_{\alpha}}^{2}}\tan^{2}\beta
   \sum_{i,j=1}^{3}T_{D}^{i3}T_{D}^{j1^{*}}
   \Biggl[\frac{U^{A}_{\alpha1}U^{A}_{\alpha1}}{m_{b}}\delta_{ij}
   P_{1}(x_{\tilde{b}_{i-1}\tilde{g}}) \nonumber\\
&& \hspace*{6cm}
   +\frac{U^{A}_{\alpha1}T_{3}^{\alpha ji}}{m_{\widetilde{b}_{j-1}}^{2}}
   F_{1}(x_{\tilde{b}_{i-1}\tilde{b}_{j-1}},x_{\tilde{g}\tilde{b}_{j-1}})\Biggl].
\end{eqnarray}
\normalsize
In the above expressions the constants and functions are defined by
\small
\begin{eqnarray}
&& N_{j}'=\frac{1}{3}Z_{N}^{1j^{*}}\sin\theta_{w}-Z_{N}^{2j^{*}} \cos\theta_{w},~~
          N_{j}''=-Z_{N}^{1j}\sin\theta_{w}+Z_{N}^{2j}\cos\theta_{w}\\
&& R_{\alpha lj}=-\frac{g_{2}}{\sqrt{2}}(U^{A}_{\alpha1}Z_{-}^{2l}Z_{+}^{1j}
   +U^{A}_{\alpha2}Z_{-}^{1l}Z_{+}^{2j})-\frac{\lambda}{\sqrt{2}}U^{A}_{\alpha3}Z_{-}^{2l}Z_{+}^{2j} \\
&& Q_{alj}=\frac{g_{2}}{\sqrt{2}}(U^{H}_{a1}Z_{-}^{2l}Z_{+}^{1j}+U^{H}_{a2}Z_{-}^{1l}Z_{+}^{2j})
    -\frac{\lambda}{\sqrt{2}}U^{H}_{a3}Z_{-}^{2l}Z_{+}^{2j} \\
&& R_{\alpha jk}^{L''}=-R_{\alpha jk}^{R''^{*}}=
    -\frac{1}{2}\{-U^{A}_{\alpha2}[\frac{g_{2}}{\cos\theta_{w}}
  (Z_{N}^{4k}N_{j}^{''}+Z_{N}^{4j}N_{k}^{''})
  +\sqrt{2}\lambda(Z_{N}^{5j}Z_{N}^{3k}+Z_{N}^{3j}Z_{N}^{5k})]\nonumber \\
&& \hspace*{2.5cm}
   +U^{A}_{\alpha1}[\frac{g_{2}}{\cos\theta_{w}}(Z_{N}^{3k}N_{j}^{''}+Z_{N}^{3j}N_{k}^{''})
   -\sqrt{2}\lambda(Z_{N}^{5j}Z_{N}^{4k}+Z_{N}^{4j}Z_{N}^{5k})]\} \nonumber\\
&& \hspace*{2.5cm}
   -\sqrt{2}\kappa U^{A}_{\alpha3}(Z_{N}^{5j}Z_{N}^{5k}+Z_{N}^{5j}Z_{N}^{5k}) \\
&& Q_{ajk}^{L''}=Q_{\alpha jk}^{R''}=
   \frac{1}{2}\{U^{H}_{a2}[\frac{-g_{2}}{\cos\theta_{w}}(Z_{N}^{4k}N_{j}^{''}
   +Z_{N}^{4j}N_{k}^{''})
   +\sqrt{2}\lambda(Z_{N}^{5j}Z_{N}^{3k}+Z_{N}^{3j}Z_{N}^{5k})]\nonumber \\
&& \hspace*{2.5cm}
   +U^{H}_{a1}[\frac{g_{2}}{\cos\theta_{w}}(Z_{N}^{3k}N_{j}^{''}+Z_{N}^{3j}N_{k}^{''})
   +\sqrt{2}\lambda(Z_{N}^{5j}Z_{N}^{4k}+Z_{N}^{4j}Z_{N}^{5k})]\} \nonumber\\
&& \hspace*{2.5cm}
   -\sqrt{2}\kappa U^{H}_{a3}(Z_{N}^{5j}Z_{N}^{5k}+Z_{N}^{5j}Z_{N}^{5k})\\
&& \Gamma_{1}(i,k,j,l)=(T_{U}^{i2}T_{U}^{k2^{*}}-\delta_{i1}\delta_{k1})Z_{+}^{1l^{*}}Z_{-}^{2j^{*}}
   -\frac{m_{t}}{\sqrt{2}m_{W}\sin\beta}T_{U}^{i3}T_{U}^{k2^{*}}Z_{+}^{2l^{*}}Z_{-}^{2j^{*}}\\
&& \Gamma_{2}(i,j,k')=\sqrt{2}\tan\theta_{w}Z_{N}^{1j^{*}}T_{D}^{i1^{*}}T_{D}^{k'3}
    +\frac{m_{b}}{\sqrt{2}m_{W}\cos\beta}Z_{N}^{3j^{*}}T_{D}^{i1^{*}}T_{D}^{k'2} \\
&& \Gamma_{6}(i,j)=\frac{m_{l}\sin\theta_{w}}{3m_{b}\cos\theta_{w}}
        Z_{N}^{1j^{*}}T_{D}^{i1^{*}}T_{D}^{i3} \\
&& \Gamma_{7}(i,j,k)=\frac{1}{2\cos\theta_{w}}Z_{N}^{3j}T_{D}^{i1^{*}}T_{D}^{i2}
    [Z_{L}^{Im^{*}}Z_{L}^{Im}(Z_{N}^{1j^{*}}\sin\theta_{w}+Z_{N}^{2j^{*}}
    \cos\theta_{w})Z_{N}^{3k^{*}} \nonumber\\
&& \hspace*{2cm}
    -2\sin\theta_{w}Z_{L}^{(I+3)m^{*}}Z_{L}^{(I+3)m}Z_{N}^{3j^{*}}Z_{N}^{1k^{*}}] \\
&& T_{1}^{\alpha ik}=A_{3}T_{U}^{i3^{*}}T_{U}^{k2}-A_{3}^{*}T_{U}^{i2^{*}}T_{U}^{k3},~~
     T_{3}^{\alpha ik}=A_{1}T_{D}^{k2^{*}}T_{D}^{i3}-A_{1}^{*}T_{D}^{k3^{*}}T_{D}^{i2} \\
&& T_{2}^{aik}=-\frac{m_{t}}{2m_{W}}[2m_{t}U^{H}_{a2}(T_{U}^{i2^{*}}T_{U}^{k2}
     +T_{U}^{i3^{*}}T_{U}^{k3})
      +(A_{4}T_{U}^{i3^{*}}T_{U}^{k2}+A_{4}^{*}T_{U}^{i2^{*}}T_{U}^{k3})]\non
  &&  ~~~~~~~~~~
   +A_{5}(T_{U}^{i1^{*}}T_{U}^{k1}+T_{U}^{i2^{*}}T_{U}^{k2})+A_{6}T_{U}^{i3^{*}}T_{U}^{k3} \\
&& T_{4}^{aik}=-2m_{t}U^{H}_{a1}(T_{D}^{i2}T_{D}^{k2^{*}}+T_{D}^{i3}T_{D}^{k3^{*}})
    +(A_{2}T_{D}^{i3}T_{U}^{k2^{*}}+A_{2}^{*}T_{U}^{i2}T_{U}^{k3^{*}})\\
&& T_{5}^{aik}=A_{7}T_{D}^{i3}T_{U}^{k3^{*}}-A_{8}(T_{D}^{i2}T_{D}^{k2^{*}}+T_{D}^{i1}T_{D}^{k1^{*}}) \\
&& A_{1}=\frac{\lambda}{\sqrt{2}}(v_{u}U^{A}_{\alpha3}+sU^{A}_{\alpha2})-A_{D}U^{A}_{\alpha1}, ~~
   A_{2}=\frac{\lambda}{\sqrt{2}}(v_{u}U^{H}_{a3}+sU^{H}_{a2})+A_{D}U^{H}_{a1} \\
&& A_{3}=\frac{\lambda}{\sqrt{2}}(v_{d}U^{A}_{\alpha3}+sU^{A}_{\alpha1})-A_{U}U^{A}_{\alpha2}, ~~
   A_{4}=\frac{\lambda}{\sqrt{2}}(v_{d}U^{H}_{a3}+sU^{H}_{a1})+A_{U}U^{H}_{a2} \\
&& A_{5}=\frac{m_{z}}{2\cos\theta_{w}}(1-\frac{4}{3}\sin^{2}\theta_{w})U^{H}_{a2}, ~~
   A_{6}=\frac{2}{3}m_{W}\tan^{2}\theta_{w}U^{H}_{a2}\\
&&  A_{7}=m_{W}e_{d}\tan^{2}\theta_{w}U^{H}_{a2}, ~~
   A_{8}=\frac{m_{z}}{2\cos\theta_{w}}(1+2e_{d}\sin\theta_{w})U^{H}_{a2}\\
&& P_{1}(x)=\frac{x\ln x}{x-1},~~
    F_{1}(x,y)=\frac{1}{x-y}\left(\frac{x\ln x}{x-1}-\frac{y\ln y}{y-1}\right), \\
&&  F_{2}(x,y)=\frac{1}{x-y}\left(\frac{x^{2}\ln x}{x-1}-\frac{y^{2}\ln y}{y-1}\right) \\
&& F_{3}(x,y,z)=\frac{x\ln x}{(x-1)(x-y)(x-z)}+(x\leftrightarrow y)+(x\leftrightarrow z),~ \\
&& F_{4}(x,y,z)=\frac{x^{2}\ln x}{(x-1)(x-y)(x-z)}+(x\leftrightarrow y)+(x\leftrightarrow z).
\label{fun}
\end{eqnarray}


\normalsize

\begin{thebibliography}{99}

\bibitem{reviewMSSM} For recent reviews, see, V. Barger, P. Langacker, H.-S. Lee, G. Shaughnessy,
                                 \PRD73,(2006) 115010.
                     V. Barger, P. Langacker, G. Shaughnessy, hep-ph/0611112.

\bibitem{hierarchy} R. Dermisek, J. F. Gunion,
                           \PRL95,(2005) 041801;  \PRD73,(2006) 111701.

\bibitem{He-HyperCP} X.-G. He, J. Tandean, G. Valencia, \PRL98, (2007) 081802;
                     \PRD74, (2006) 115015.

\bibitem{hiller} G. Hiller,  \PRD70,(2004) 034018.

\bibitem{Ellwanger} F. Domingo and U. Ellwanger, arXiv: 0710.3714.
                    For similar studies in the MSSM, see,
             M. Carena,  {\it et al.},  \PRD74, 015009 (2006);  \PRD76, 035004 (2007);
                     G. Isidori and P. Paradisi, \PLB639, 499 (2006);
                     J. Ellis,   {\it et al.}, arXiv: 0706.0652 [hep-ph];
                     M. Blanke,  {\it et al.}, JHEP 0610, 003 (2006);
                     P. Ball and R. Fleischer,  Eur. Phys. J. C {\bf 48}, 413 (2006);
                     M. Blanke and A. J. Buras, JHEP 0705, 061 (2007);
                     A. Freitas, E. Gasser and U. Haisch, \PRD76, 014016 (2007);
                     G. Isidori,  {\it et al.}, \PRD75, 115019 (2007);
                     W. Altmannshofer, A. J. Buras and D. Guadagnoli, hep-ph/0703200.

\bibitem{hurth}  For a review, see, T. Hurth, Rev. Mod. Phys. {\bf 75}, 1159 (2003).

\bibitem{Ghinculov04} A. Ghinculov, T. Turth, G. Isidori and Y.-P. Yao, \NPB685, 351 (2004);
              F. M. Borzumati and C. Greub, \PRD58, 074004 (1998);
                      K.~Chetykin, M.~Misiak, and M.~M\"{u}nz, \PLB400, 206 (1997);
              C.~Greub and T.~Hurth, \PRD56, 2934 (1997);
                     J.~L.~Hewett,  \PRD53, 4964 (1996);
                      M. Misiak, \NPB393, 23 (1993);

\bibitem{Dai97}  Y.~B.~Dai, C.~S.~Huang and H.~W.~Huang, \PLB390, 257 (1997);

\bibitem{Huang99}  C.~S.~Huang, W. Liao, Q. Yan, \PRD59, 011701 (1999);
                       C.~S.~Huang, S. H. Zhu, \PRD61, 015011 (2000);
                       C.~S.~Huang, W. Liao, Q. Yan, S. H. Zhu, \PRD63, 114021 (2001);
                          Eur. Phys. J. C {\bf 25}, 103 (2002);
        S.~R.~Choudury, N.~Gaur, \PLB451, 86 (1999).

\bibitem{Aliev97}
  T.~Goto, Y.~Okada, Y.~Shinizu and M.~Tanada, \PRD55, 4273 (1997);
  T.~M.~Aliev, A.~\"{O}zpineci and M.~Savci, \PRD55, 7059 (1997);
  T.~M.~Aliev, N.~K.~Pak and M.~Savci, \PLB424, 175 (1998).

\bibitem{Iltan00} E.~O.~Iltan and G.~Turan, \PRD61, 034010 (2000).

\bibitem{Xiong01} Z.~Xiong and J.~M.~Yang,  \NPB602, 289 (2001);

\bibitem{other}  C. Bobeth, T. Ewerth, F. Kruger, J. Urban, \PRD66, 074021 (2002);
                 G. Erkol, G. Turan, JHEP 0202, 015 (2002); \NPB635, 286 (2002);
                 S. R. Choudhury, N. Gaur, \PRD66, 094015 (2002);
                 J. K. Mizukoshi, X. Tata, Y. Wang, \PRD66, 115003  (2002);
                 Y. B. Dai, C. S. Huang, J. T. Li, W. J. Li, \PRD67, 096007  (2003);
                 C. S. Huang, X. H. Wu, \NPB657, 304 (2003);
                 Y. Wang, D. Atwood, \PRD68, 094016 (2003);
                 A. S. Cornell, N. Gaur,  JHEP 0309, 030 (2003);
                 U. O. Yilmaz, B. B. Sirvanli, G. Turan, Eur. Phys. J. C {\bf 30}, 197 (2003);
                 T. Feldmann, J. Matias, JHEP 0301, 074 (2003);
                 T. Ibrahim, P. Nath, \PRD67, 016005  (2003);
                 P.H. Chankowski, L. Slawianowska, Eur. Phys. J. C {\bf 33}, 123 (2004);
                 S. R. Choudhury, A. S. Cornell, N. Gaur, G. C. Joshi, \PRD69, 054018 (2004);
                 T. M. Aliev, V. Bashiry, M. Savci, JHEP 0405, 037 (2004); \PRD73, 034013 (2006);
                  H. Acar, G. Turan, Acta Phys. Polon. B35, 687 (2004) [hep-ph/0410125];
           U. O. Yilmaz, B. B. Sirvanli, G. Turan, \NPB692, 249 (2004);
                   T.-F. Feng, X.-Q. Li, J. Maalampi, \PRD73, 035011 (2006);
                   Z.~Xiong and J.~M.~Yang,  \NPB628, 193 (2002);
                   Y.~Su, \PRD56, 335 (1997).

\bibitem{Miller:2003ay} D.~J.~Miller, R.~Nevzorov and P.~M.~Zerwas, \NPB681, 3 (2004);

\bibitem{Hikasa87} K.~I.~Hikasa and M.~Kobayashi, \PRD36, 724 (1987).

\bibitem{nmhdecay} U. Ellwanger, J. F. Gunion, C. Hugonie, JHEP {\bf0502}:066(2005).

\bibitem{Lunghi99} E.~Lunghi, A.~Masiero,~I.~Scimemi,~L.~Silvestrini, \NPB568, 120 (2000).

\bibitem{Bertolini91}
         S.~Bertolini,~F.~Borzumati,~A.~Masiero and G.~Ridolfi, \NPB353, 591 (1991);
        A.~J.~Buras and M.~M\"unz,  \PRD52, 186 (1995);
        P. Cho, M. Misiak and D. Wyler, \PRD54, (1996) 3329;
        M.~Ciuchini,~G.~Degrassi,~P.~Gambino and G.~F.~Giudice, \NPB527, 21 (1998); \NPB534,3 (1998).

\bibitem{cut_ex} J.Kaneko, {\it et al.}, \PRL90, 021801 (2003).

\bibitem{HFAG} E. Barberio, {\it et al.},  Heavy Flavor Averaging Group (HFAG), hep-ex/0603003
               (webpage: www.slac.stanford.edu/xorg/hfag).

\bibitem{PDG}  W.M. Yao {\it et al.}, J. Phys. G {\bf 33}, 1 (2006).
\bibitem{Ali1996}
   A.~Ali, G.~Hiller, L.~T.~Handoko and T.~Morozumi,
   Phys.\ Rev.\  D {\bf 55}, 4105 (1997).
\bibitem{xbsmm} Z.~Xiong and J.~M.~Yang, \PLB546, (2002) 221
\end{thebibliography}
\end{document}